\documentclass[12pt]{article}

\usepackage{arxiv}
\renewcommand{\headeright}{}  
\renewcommand{\undertitle}{}  
\usepackage{amsmath, amssymb, amsfonts, amsthm}
\usepackage{graphicx}
\usepackage{booktabs}
\usepackage{multirow}
\usepackage{arydshln}
\usepackage{multirow}
\usepackage{algorithm}
\usepackage{algpseudocode}
\usepackage{multirow}
\usepackage{subcaption,caption}
\usepackage{tikz}
\usetikzlibrary{matrix,decorations.pathreplacing, calc, positioning}

\usepackage[utf8]{inputenc} 
\usepackage[T1]{fontenc}    
\usepackage{hyperref}       
\usepackage{url}            
\usepackage{booktabs}       
\usepackage{amsfonts}       
\usepackage{nicefrac}       
\usepackage{microtype}      
\usepackage{lipsum}		
\usepackage{graphicx}
\usepackage{natbib}
\usepackage{doi}

\title{Quantifying and Attributing Submodel Uncertainty in Stochastic Simulation Models and Digital Twins}

\author{%
Mohammadmahdi Ghasemloo\\
Department of Industrial and Systems Engineering\\
Texas A\&M University\\
College Station, TX, USA \\
\texttt{mohammad\_ghasemloo@tamu.edu} \\
\And
David J.~Eckman \\
Department of Industrial and Systems Engineering\\
Texas A\&M University\\
College Station, TX, USA \\
\texttt{eckman@tamu.edu} \\
\And
Yaxian Li \\
Intuit AI\\
Mountain View, CA, USA \\
\texttt{yaxian\_li@intuit.com} \\
}

\date{}

\begin{document}
\maketitle

\begin{abstract}
	Stochastic simulation is widely used to study complex systems composed of various interconnected subprocesses, such as input processes, routing and control logic, optimization routines, and data-driven decision modules. In practice, these subprocesses may be inherently unknown or too computationally intensive to directly embed in the simulation model. Replacing these elements with estimated or learned approximations introduces a form of epistemic uncertainty that we refer to as \emph{submodel uncertainty}. This paper investigates how submodel uncertainty affects the estimation of system performance metrics.
We develop a framework for quantifying submodel uncertainty in stochastic simulation models and extend the framework to digital-twin settings, where simulation experiments are repeatedly conducted with the model initialized from observed system states. Building on approaches from input uncertainty analysis, we leverage bootstrapping and Bayesian model averaging to construct quantile-based confidence or credible intervals for key performance indicators. We propose a tree-based method that decomposes total output variability and attributes uncertainty to individual submodels in the form of importance scores.
The proposed framework is model-agnostic and accommodates both parametric and nonparametric submodels under frequentist and Bayesian modeling paradigms. A synthetic numerical experiment and a more realistic digital-twin simulation of a contact center illustrate the importance of understanding how and how much individual submodels contribute to overall uncertainty.
\end{abstract}

\keywords{Uncertainty Quantification, Digital Twin, Stochastic Simulation, AI-driven Simulation}

\section{Introduction}
\label{sec:intro}

Many stochastic systems studied in operations research and management science can be viewed as collections of interconnected subprocesses that each represent a specific aspect of system behavior such as demand generation, service dynamics, routing logic, or optimization-based decision making. When such systems are studied using stochastic simulation, the modeler must determine how to model each subprocess.
In many cases, the true subprocess cannot be accessed or fully understood and instead can only be observed in the real world, such as when observing customer arrivals in a service system.
In other cases, the true subprocess may be accessible, such as when it represents some operational decision made according to some policy or sophisticated solution method, e.g., optimization, but directly embedding the subprocess within the simulation may be impractical due to concerns about privacy, latency, computational cost, or data accessibility. For these reasons, the modeler may, out of necessity or choice, replace certain subprocesses with approximate representations, such as generative models, simplified decision rules, heuristics, or optimization proxies trained on historical data.
These approximations, which we henceforth refer to as \textit{submodels}, may take several forms. In the case of modeling stochastic inputs, such as customer demand or service times, probability distributions are fitted to real-world data. In other cases, a submodel may be a function mapping subsystem inputs to outputs trained on input-output data from that subprocess using supervised learning techniques. 
In both cases, submodel selection can involve trade-offs between accuracy, interpretability, and computational cost. 

The use of submodels within a stochastic simulation model introduces errors that propagate through the model to its outputs.
Understanding how these errors propagate is critical for quantifying uncertainty in estimators of key performance indicators (KPIs) and directing efforts to reduce uncertainty. 
While prior work in the simulation literature has extensively studied \emph{input uncertainty} (IU) \citep{barton2022input, lam2022subsampling, nelson2021foundations}---the epistemic uncertainty due to estimating stochastic input models from limited data---less attention has been given to uncertainty originating from other types of submodels that influence the internal dynamics of the simulated system, including the decision logic, event ordering, and system state. The errors introduced by both types of submodels can influence simulation behavior in complex and path-dependent ways.
Along these lines, \cite{ghasemloo2025quantifying} adopt methods from IU to study uncertainty arising from embedding machine learning (ML) surrogates of decision-support systems (DSSs) within simulation models. 

This paper builds on the ideas of \cite{ghasemloo2025quantifying} to introduce a unifying framework that encompasses many potential sources of epistemic uncertainty in stochastic simulation models. We collectively refer to this more general form of epistemic uncertainty as \emph{submodel uncertainty}.
This paper presents general-purpose methods for systematically quantifying submodel uncertainty to support and enhance operational decision making.
More specifically, we employ bootstrapping and Bayesian model averaging (BMA) to generate plausible submodels that drive a designed simulation experiment. We leverage design of experiments (DOE), specifically stacked Latin hypercube (LH) designs, to more efficiently explore the space of submodel instances when studying systems with multiple submodels.
The experiment results are then used to construct quantile-based confidence or credible intervals (CIs) that account for both aleatoric and epistemic uncertainty. 
We propose a tree-based method that provides importance scores quantifying how the overall uncertainty can be decomposed into aleatoric and epistemic terms and how the epistemic uncertainty can be further attributed to individual submodels. These importance scores enable practitioners to identify the most influential submodels and thereby prioritize further efforts to reduce overall uncertainty by, for example, acquiring additional training data or refining their modeling specifications.
The framework's strength is its generality, as it can accommodate both parametric and non-parametric submodels under either the frequentist or Bayesian modeling paradigms. Moreover, whereas some existing approaches for analyzing IU assume that the input models are independent of each other \citep{he2024introductory}, we make no such assumption about the submodels. 

Submodel uncertainty is also highly relevant in digital-twin settings, wherein a simulation model is enhanced by integrating real-time or periodically observed system state data to maintain a synchronized virtual representation of the physical system \citep{pylianidis2022simulation, huang2021survey,  taylor2023enhancing}. Digital twins are used for monitoring, forecasting, and evaluating alternative operational strategies in real time. Recent work has emphasized the central role of uncertainty quantification (UQ) for reliable prediction and decision support. For example, \cite{yang2022dt_design_uncertainty} develop a non-intrusive sensitivity-metrics toolbox for integrating black-box digital twins into the design process in the presence of uncertain data and evolving performance indicators. 

In digital-twin settings, simulation models are repeatedly initialized from observed system states and commonly rely on submodels—such as forecasting tools, routing models, or optimization proxies—to emulate future system behavior. For more discussion on the increasing use of ML submodels within digital twins, see
\citep{thelen2022comprehensive}.
We extend our framework to the digital-twin setting by accommodating cases in which the system state evolves over time and the submodels may be time dependent.
In particular, we apply our tree-based method on each state and aggregate the importance scores across states to quantify and compare the relative contributions of different submodels to uncertainty over time. We also propose a method for estimating a form of state-average bias when data on historical system states and KPIs are available. In summary, our approach provides a unified and scalable framework for uncertainty quantification and decomposition that supports reliable decision making in offline simulation and online digital-twin applications.

The remainder of the paper is organized as follows. Section~\ref{sec:submodel} formally defines submodels in simulated systems and categorizes them based on their functional forms. Section~\ref{sec:submode_uncertainty} mathematically introduces submodel uncertainty. Sections~\ref{sec:resample} and \ref{sec:uq} describe resampling-based methods for UQ and tree-based methods that attribute uncertainty to individual submodels along with a numerical example. Submodel uncertainty is extended to the digital-twin setting in Section~\ref{sec:digial_twin} and demonstrated through a contact-center example. Concluding remarks and future research directions are given in Section~\ref{sec:conclusion}.

\section{Submodels in Simulated Systems}
\label{sec:submodel}

A stochastic simulation model can be viewed as a mapping $g$ that takes as input a collection of user-specified variables $\mathbf{X}$ and random primitives $\xi$ and returns a response $\mathbf{Y}$, i.e., 
\begin{equation}
\label{eq:general_model}
\mathbf{Y} = g(\mathbf{X}, \xi).
\end{equation}
Simulation models typically consist of multiple interconnected subprocesses, each of which emulates a specific aspect of system behavior. Various practical limitations often make it infeasible to incorporate the real-world subprocess directly in the simulation model. First, the true subprocess may not be fully accessible. For example, input processes governing customer arrivals to a service system are inherently unavailable.
Second, even when the true subprocess is accessible, data intensity and latency constraints often render it infeasible to make repeated calls to the true subprocess from within the simulation model. 
Third, in many real-world settings, decision modules, such as those that support routing, dispatch, and allocation decisions, are shared across real-time applications. Directly invoking these modules within a simulation experiment may therefore interfere with live operations, introduce contention for computational resources, or expose sensitive information. These issues can generally be resolved by approximating or replacing the real-world subprocesses with submodels.
For a more detailed discussion of the motivations for replacing DSSs with ML surrogate models, see \cite{ghasemloo2025quantifying}. 

The submodels we investigate in this paper can generally be expressed in the form of \eqref{eq:general_model}. We categorize submodels into three classes depending on whether their mapping $g$ features only $\mathbf{X}$, only $\xi$, or both $\mathbf{X}$ and $\xi$.

\subsection{Deterministic Submodels}
We call submodels for which $\mathbf{Y} = g(\mathbf{X})$ \textit{deterministic submodels}, because their output is a deterministic function of the input variables. Deterministic submodels do not introduce any aleatoric uncertainty to the simulation. Roughly speaking, deterministic submodels can be thought of as rules-based logic in the simulation that maps the system state $\mathbf{X}$ to a deterministic decision or action $\mathbf{Y}$. 
A deterministic submodel $g$ is trained on data consisting of paired observations $(\mathbf{X}, \mathbf{Y})$ and then embedded within the simulation model. The mapping $g$ may correspond to an optimization proxy \citep{chen2024real} that replaces an optimization solver and is invoked within the simulation whenever an instance of the corresponding optimization problem needs to be solved. In this case, $\mathbf{X}$ denotes the set of parameters describing the problem instance and $\mathbf{Y}$ is the resulting optimal solution, which is then used to advance the simulation.
Alternatively, $g$ may approximate the actions of a reinforcement learning (RL) \citep{sutton1998reinforcement} or otherwise intelligent agent, where $\mathbf{X}$ is the state of the system and $\mathbf{Y}$ is the associated action. Inverse reinforcement learning (IRL) \citep{arora2021survey} can be used to learn a reward function from historical state-action observations and subsequently train a policy. The submodel $g$ might then be the table in tabular learning or be parameterized by a deep neural network \citep{wulfmeier2015maximum} that determines the optimal policy. The DSSs studied in \cite{ghasemloo2025quantifying} are examples of deterministic submodels.

\subsection{Unconditional Stochastic Submodels}

We refer to submodels that inject aleatoric uncertainty into the simulation model and are not influenced by external factors as \textit{unconditional stochastic submodels}. These submodels are of the form $\mathbf{Y} = g(\xi)$,
where $\xi$ represents random primitives, $\mathbf{Y}$ is a random quantity, and $g$ can be parametric or nonparametric. Training data for this class of submodels consist solely of observations of $\mathbf{Y}$. Well-known examples of unconditional stochastic submodels arise in random variate generation, where the function $g$ could be an inverse cumulative distribution function that transforms a pseudo-random number $\xi \sim \mathcal{U}(0,1)$ into a random variate $\mathbf{Y}$ or could represent an acceptance-rejection algorithm given access to a sequence of pseudo-random numbers \citep{devroye2006nonuniform}. In more modern generative modeling approaches, such as neural input modeling (NIM) \citep{cen2023nim}, the mapping $g$ may be parameterized by a neural network, for example, within a variational autoencoder (VAE), while the latent variable $\xi$ is typically assumed to follow a standard normal distribution. Alternatively, $g$ may be represented by neural architectures that incorporate recurrent components, such as long short-term memory (LSTM) layers, to capture temporal dependencies in the input process \citep{cen2023nim}.

\subsection{Conditional Stochastic Submodels}

We refer to submodels of the general form
$\mathbf{Y}= g(\mathbf{X}, \xi)$ as \textit{conditional stochastic submodels}. 
As an example, $g$ may be a generative metamodel (surrogate) of a stochastic simulation model embedded within a larger simulation framework. In this case, $\mathbf{X}$ may include the system state used to execute the submodel, while $\xi$ is generated from the stochastic input models that govern the simulation submodel's aleatoric uncertainty.
Building upon a previous example of deterministic submodels, $g$ may instead represent a \textit{stochastic} optimization proxy \citep{alcantara2025quantile} that replaces a stochastic optimization solver within the simulation model.
Here, $\mathbf{X}$ is again the set of parameters describing the problem instance and $\xi$ captures either inherent stochasticity in the optimization algorithm or that emanating from a stochastic submodel (e.g., a simulation submodel) invoked by the solver in the course of its run. Random variate generators that depend on contextual information $\mathbf{X}$, such as conditional variational autoencoders (CVAEs) \citep{cen2023nim}, also fall into this category.

\subsection{Supply Chain Example}
We now provide an illustrative example of a supply chain simulation model featuring the three types of submodels. Consider a multi-echelon supply chain in which products must be dynamically allocated from factories to distribution centers and retail locations based on incoming demand information and the current state of inventories and transportation resources. Decisions made at each epoch influence future inventory levels, production schedules, and transportation availability across the network.
A stochastic simulation model of this system may contain several submodels. First, customer demand arrivals and their associated order quantities can be represented through unconditional stochastic submodels that generate demand realizations and serve as inputs to the simulation model \citep{simchi2007designing}.
Second, allocation decisions regarding where to send each product across the network may be carried out by repeatedly invoking smaller embedded simulation models that evaluate mean completion times \citep{simio}. Or, if such simulation submodels are too computationally expensive to embed, they could be replaced with cheaper conditional stochastic submodels, such as CVAEs, that generate synthetic outputs of the simulation submodel.
Finally, within each factory, production scheduling and sequencing decisions may be made by RL agents trained to optimize operational performance over time. When the true RL policy is inaccessible or too computationally expensive to deploy online, its behavior can be approximated by a deterministic submodel that learns the reward function using IRL. 

\section{Submodel Uncertainty}
\label{sec:submode_uncertainty}

Let \( \mathbf{S}^c= (\mathbf{S}^c_1, \mathbf{S}^c_2, \ldots, \mathbf{S}^c_L) \) represent a collection of $L$ true subprocesses that govern the real-world system. When hypothetically driving a stochastic simulation model with the true subprocesses, the (random) output on a given replication \( j = 1, 2, \ldots \) can be expressed as
\[
Y_j(\mathbf{S}^c) = \mu(\mathbf{S}^c) + \varepsilon_j(\mathbf{S}^c),
\]
where \( \mu(\mathbf{S}^c) \equiv \mathbb{E}[Y_j(\mathbf{S}^c)] \) represents the expected output, and \( \varepsilon_j( \mathbf{S}^c) \) is a mean-zero random variable capturing the aleatoric uncertainty in the output.
In this paper, we study the estimation of a single expected simulation output, hence $Y_j(\mathbf{S}^c)$ is a scalar.

A standard approach to estimating \( \mu(\mathbf{S}^c) \) is to obtain estimated submodels \( \hat{\mathbf{S}} \) from real-world data, run $n$ independent replications using $\hat{\mathbf{S}}$, and compute
\[
\bar{Y}(\hat{\mathbf{S}}) = \frac{1}{n} \sum_{j=1}^n Y_j(\hat{\mathbf{S}}).
\]
A confidence interval for \( \mu({\mathbf{S}^c}) \) can then be constructed as \( \bar{Y}( \hat{\mathbf{S}}) \pm t_{\alpha/2, n-1} \cdot s / \sqrt{n} \), where \( s \) is the sample standard deviation of $Y_1(\hat{\mathbf{S}}), Y_2(\hat{\mathbf{S}}), \ldots, Y_n(\hat{\mathbf{S}})$, and \( t_{\alpha/2, n-1} \) is the $1-\alpha/2$ quantile of a Student's $t$-distribution with $n-1$ degrees of freedom.
This confidence interval, however, overlooks the epistemic uncertainty from estimating \( \mathbf{S}^c \) using  \( \hat{\mathbf{S}} \) and therefore may fail to cover $\mu(\mathbf{S}^c)$ with the prescribed confidence.
This undercoverage can be explained by the observation that replacing \( \mathbf{S}^c \) with $\hat{\mathbf{S}}$ can affect both the variance and bias of the estimator $\bar{Y}( \hat{\mathbf{S}})$.

Understanding the variance of $\bar{Y}(\hat{\mathbf{S}})$ is crucial for building an asymptotically valid confidence interval  or credible interval for $\mu(\mathbf{S}^c)$. Applying the law of total variance, the variance of $\bar{Y}(\hat{\mathbf{S}})$---where $\hat{\mathbf{S}}$ is treated as random---can be decomposed as 
$$
\text{Var}(\bar{Y}(\hat{\mathbf{S}})) = \text{Var}(\mathbb{E}[\bar{Y}(\hat{\mathbf{S}}) \mid \hat{\mathbf{S}}]) + \mathbb{E}[\text{Var}(\bar{Y}(\hat{\mathbf{S}}) \mid\hat{\mathbf{S}})] \equiv \sigma_{\mathrm{SU}}^2 + \sigma_{\mathrm{MC}}^2.  
$$
This decomposition separates the aleatoric uncertainty (i.e., the simulation noise) \( \sigma_{\mathrm{MC}}^2 \), from the epistemic uncertainty introduced by the use of submodels, \( \sigma_{\mathrm{SU}}^2 \). This additive decomposition of the total uncertainty into epistemic and aleatoric uncertainty has been used extensively in the IU literature \citep{barton2022input}.
We further attribute $\sigma^2_{\mathrm{SU}}$ across submodels in Section \ref{sec:uq}.

The presence of estimated submodels also affects the bias of the estimator:
\[
\text{Bias} = \mathbb{E}[\mu(\hat{\mathbf{S}})] - \mu(\mathbf{S}^c),
\]
where the expectation in the first term is taken over both $\hat{\mathbf{S}}$ and replications of the simulation model.
Although bias estimation is not the primary focus of this paper, it nonetheless can play an important role in how well the resulting CIs cover $\mu(\mathbf{S}^c)$. Bias has received less attention in the IU literature, partly because it is difficult to estimate and is expected to vanish faster than the variance as the size of the training data increases asymptotically \citep{morgan2019detecting}. In Section 6, we propose a method to estimate a related form of bias when outputs from the real-world system are available. 

Our framework for studying the effects of submodel uncertainty on KPI estimation entails obtaining random instances of each  submodel and using them to drive the simulation model in a designed experiment. The next two sections elaborate on how to generate these instances, design the experiment, and analyze the results.

\section{(Re)Sampling Submodels}
\label{sec:resample}
In this section, we discuss ways to sample from the space of submodels when they are formulated under the frequentist and Bayesian statistical philosophies.

\subsection{Frequentist Submodels and Bootstrapping}
\label{subsec:bootstrap}

In the frequentist paradigm, a submodel is assumed to have a fixed but unknown structure or parameterization. Parametric submodels are typically constructed by first selecting a functional or distributional form based on domain knowledge and then estimating parameters from data. Uncertainty arises from limited data and submodel misspecification and is commonly assessed through resampling-based techniques such as bootstrapping \citep{efron1994introduction}.
Bootstrapping is a non-parametric resampling technique that involves resampling observations \textit{with replacement} to create multiple bootstrapped datasets and has been extensively applied to quantify IU \citep{barton2022input}. A straightforward way to leverage bootstrapping for studying submodel uncertainty is to resample the training data with replacement and generate plausible instances of the submodel by successively fitting it to the bootstrapped datasets \citep{tang2024consistency}. 
Although bootstrapping does not impose strong distributional assumptions on the data, it can be challenging to apply in certain contexts, e.g., time series data, which could arise, for example, if the submodel were a forecasting tool. Specialized techniques, such as moving block bootstrap \citep{lahiri2003resampling} or stationary bootstrap \citep{politis1994stationary}, have been developed to address this case, but they can be harder to implement.
Bootstrapping also tends to multiply the required computational effort by the number of bootstrap instances. In certain IU settings, bootstrapping is relatively inexpensive because it entails fitting probability distributions to data, which is typically fast. In contrast, when bootstrapping is used to quantify, for instance, a deterministic or conditional stochastic submodel's uncertainty, the computational burden can be higher because a supervised learning model needs to be trained on each bootstrapped dataset. Complex ML models, such as deep neural networks, often require substantial training time, especially when the data are high dimensional or the model architecture is deep. When using bootstrapping for IU, the dominant cost is often the simulation runs themselves, whereas with submodel uncertainty the cost of repeatedly training submodels may become comparable or even dominant.

\subsection{Bayesian Submodels and Bayesian Model Averaging}
\label{subec:bayes}

Bayesian submodels explicitly consider uncertainty by treating submodel parameters as random variables. Prior distributions encode existing knowledge or beliefs and are updated using observed data to produce posterior distributions. Bayesian model averaging offers a principled framework for accounting for uncertainty in the form of a posterior distribution---obtained via Bayes' rule---which inherently captures the remaining uncertainty about the model after observing the data \citep{chick2001input}. This probabilistic formulation yields a collection of plausible submodels when sampling  from the posterior.

Bayesian models, e.g., GPs or BNNs, are equipped with posterior distributions on their predictions as functions of inputs for GPs~\citep{rasmussen2006gaussian} or parameter weights for BNNs~\citep{neal2012bayesian}.
Instead of resampling and retraining the submodel multiple times, as required in bootstrapping, Bayesian model averaging requires simply sampling from the posterior distribution of the model parameters. For example, when working with BNNs, generating new submodels requires only sampling from the posterior distribution rather than performing a full retraining process. On the other hand, sampling from GPs becomes increasingly expensive as the number of evaluation points grows, which in our setting corresponds to the number of times the submodel is invoked within a simulation replication.

\section{Uncertainty Quantification}
\label{sec:uq}
Let \( \{ \hat{\mathbf{S}}_l^{(b)} \colon b = 1, 2, \ldots, B \text{ and } l = 1, 2, \ldots, L\} \) represent \( B \) instances of each submodel generated via either bootstrapping or BMA. We assume a common $B$ across submodels, but this is not strictly necessary.
It can be prohibitively expensive to simulate all $B^{L}$ possible combinations (henceforth called configurations) of these instances, even for moderate values of $B$ and $L$. To address this challenge, we turn to design of experiments (DOE) to choose a smaller design of configurations that covers the space of submodel instances.
In this paper, we employ Latin Hypercube (LH) designs \citep{santner2018design}, which ensure that every level of every factor (in our context, every sampled instance of every submodel) appears exactly once in a design. More specifically, a LH design is constructed by sampling \textit{without replacement} from the set of $B$ instances of each submodel to form configurations until all instances have been exhausted.
This process can be repeated to create additional LH designs, which can then be stacked to form a larger design \citep{sanchez2020work}. The total number of configurations in such a stacked design is $B' = BS$, where $S$ is the number of stacks. The simulation model is then run $n$ times at each configuration.

The choices of $B$, $S$, and $n$ affect the computational cost of the experiment and the accuracy of our estimation of the distribution of $\bar{Y}(\hat{\mathbf{S}})$. 
Increasing \( B \) enhances the diversity of the sampled submodel instances, increasing \( S \) improves the coverage of the cross-product of those instances, and increasing \( n \) reduces the impact of simulation noise by averaging over more replications. Note that the number of submodels that must be fitted or retrained depends on \( B \), not \( S \).  

Our framework can also accommodate time-dependent submodels, e.g., models of time non-homogeneous arrivals. If we can assume that the behavior of the submodel is unchanging within certain periods, then the submodel can be further divided into distinct submodels---one for each period---and our approach applied as described.

We next describe how one can construct a CI for $\mu(\mathbf{S}^c)$ and use regression trees to decompose and attribute overall uncertainty to individual submodels.

\subsection{Confidence and Credible Intervals and Decomposing Uncertainty}
Suppose the sample means \( \{\bar{Y}^{\langle 1 \rangle}, \bar{Y}^{\langle 2 \rangle}, \dots, \bar{Y}^{\langle B' \rangle}\} \) are obtained from the designed experiment, where
$$
\bar{Y}^{\langle b' \rangle} = \frac{1}{n} \sum_{j=1}^{n} Y_j(\hat{\mathbf{S}}^{\langle b' \rangle}) \quad \text{for } b' = 1, 2, \ldots, B'.
$$
An approximate $100(1-\alpha)\%$ quantile-based confidence or credible interval for $\mu(\mathbf{S}^c)$ is given by
\begin{equation}
    \label{eq:conf}
    \left[ Q_{\alpha/2} \left( \bar{Y} \right), Q_{1 - \alpha/2} \left( \bar{Y} \right) \right],
\end{equation}
where \( Q_p(\bar{Y}) \) denotes the empirical $p$-quantile of the sample means.
This CI incorporates both submodel and aleatoric uncertainty. In the frequentist setting, the confidence interval is asymptotically valid as the size of the training datasets grows large, assuming that the true subprocesses belong to the chosen classes of submodels. 
Note that the CI in (\ref{eq:conf}) does not distinguish the submodels' individual contributions to overall uncertainty. Consequently, decision makers looking to reduce uncertainty---such as by increasing the number of simulation replications, collecting more real-world data, or enhancing the submodels through hyperparameter tuning or switching to a more flexible architecture---could benefit from techniques that attribute uncertainty to specific submodels.

We present a method for quantifying the contribution of different sources to overall uncertainty.
Our approach leverages \textit{variable importance} metrics from regression trees \citep{ISLR}, which assign a value to each feature based on the total reduction in the total sum of squares (TSS) resulting from splits on that feature. Our approach strongly resembles analysis of variance (ANOVA); however, whereas ANOVA separately accounts for interaction effects, tree-based importance scores absorb interactions into the main effects. Interaction contributions are typically difficult to interpret and may not correspond to actionable modeling or data-collection decisions. For example, even if a strong interaction is detected between two submodels, it may not be clear whether additional data should be collected for one submodel, the other, or both simultaneously.
Because our approach forces the interaction effects on certain submodels, we do not assume that submodels are independent, which we believe is unlikely to hold in practice. Furthermore, our approach can preserve any inherent or assumed dependence across submodels by bootstrapping the training data or sampling from the joint posterior distribution accordingly.

We first consider a regression tree trained on the dataset produced by the designed experiment, where the $L$ submodels are treated as categorical features with levels corresponding to the $B$ sampled submodel instances, e.g., $\hat{\mathbf{S}}^{(1)}_\ell, \hat{\mathbf{S}}^{(2)}_\ell , \ldots, \hat{\mathbf{S}}^{(B)}_\ell$ for $\ell=1, 2, \ldots, L$. The tree is grown using recursive binary splitting, iteratively minimizing the residual sum of squares (RSS), until each configuration is its own terminal node, at which point the RSS corresponds to the portion of the TSS that is not explained by the  submodels. The TSS is defined as
$$
\mathrm{TSS} = \sum_{b'=1}^{B'} \sum_{j=1}^n \left(Y_j^{\langle b' \rangle} - \bar{\bar{Y}}\right)^2,
\quad \text{where} \quad \bar{\bar{Y}} = \frac{1}{nB'} \sum_{b'=1}^{B'} \sum_{j=1}^n Y_j^{\langle b' \rangle},
$$
and the RSS is given by:
\[
\mathrm{RSS} = \sum_{b'=1}^{B'} \sum_{j=1}^n \left(Y_j^{\langle b' \rangle} - \hat{Y}_j^{\langle b' \rangle}\right)^2,
\]
where $\hat{Y}_j^{\langle b' \rangle}$ is the predicted value of the KPI for configuration $b'$ from the fully grown regression tree, which in our case is $\bar{Y}^{\langle b' \rangle}$.

To further understand the role of model complexity in this framework, we consider the concept of \textit{degrees of freedom} (DoF), which represents the effective number of parameters used by a model to fit data. In regression settings, DoF is often interpreted as a measure of a model’s flexibility or complexity and plays a critical role in error decomposition~\citep{ye1998gdf}.
For regression trees, a common approximation for DoF is the number of terminal (leaf) nodes~\citep{hastie2009elements}. Furthermore, the contribution of individual features to model complexity is the number of times each feature is used for splitting, where each split contributes approximately one additional DoF~\citep{mentch2020randomization}.
To attribute uncertainty to each source, we compute the reduction in TSS attributable to each submodel and normalize it by the number of times the corresponding submodel appears in the tree splits. Finally, we divide the RSS by its DoF, \( nB' - B' + 1 \), to obtain the simulation error's contribution to the overall variance.

Regression trees, are known to exhibit high variance, as small perturbations in the training data can lead to substantially different tree structures. To mitigate this variability, we adopt a bootstrap aggregation (bagging) strategy \citep{breiman2001random}. Specifically, we first generate bootstrap training datasets by resampling observations with replacement. For each bootstrap sample, a regression tree is trained and feature importance scores are computed as described above. The resulting importance scores are then averaged across bootstrap instances, in a manner analogous to the aggregation mechanism used in random forests. This aggregation can also be performed using a weighted average, where the weights are derived either from the distribution of the observed system states or from the relative magnitude of the total TSS associated with each tree. This procedure yields a set of importance estimates with reduced variance relative to those obtained from a single regression tree, without needing to train additional submodels or run more simulation replications.

\subsection{Numerical Example}
\label{sec:example_1}

In our first experiment, we study the effect of submodel uncertainty on the output of a synthetic simulation model.
We consider two independent stochastic input models $X_1 \sim \mathcal{N}(1, 0.5^2)$ and 
 $X_2 \sim \mathcal{N}(1, 2^2)$
and two deterministic transformation functions $p(x) = x^7 + x^4 + 3x^3 + \sin(x) + 4$ and $q(x) = x^3 + x^2 + 4x$.
The output of the simulation model is defined as $Y = p(X_1) + q(X_2)$
and our quantity of interest is the expected value $\mathbb{E}[Y]$, which can be derived analytically and equals $49.13$.

We examine three cases that differ in terms of the sources of epistemic uncertainty that are accounted for.
\begin{enumerate}
\item \textbf{No epistemic uncertainty:} The simulation is run using only the estimated submodels.
\item \textbf{Input uncertainty only:} UQ is performed for the input models associated with $X_1$ and $X_2$, while only the estimated submodels for $p$ and $q$ are used. This setting represents the existing practice of analyzing IU. 
\item \textbf{Submodel uncertainty:} UQ is performed for all submodels, including input models.
\end{enumerate}

\begin{figure}[htbp]
    \centering
    \includegraphics[width=1.05\linewidth]{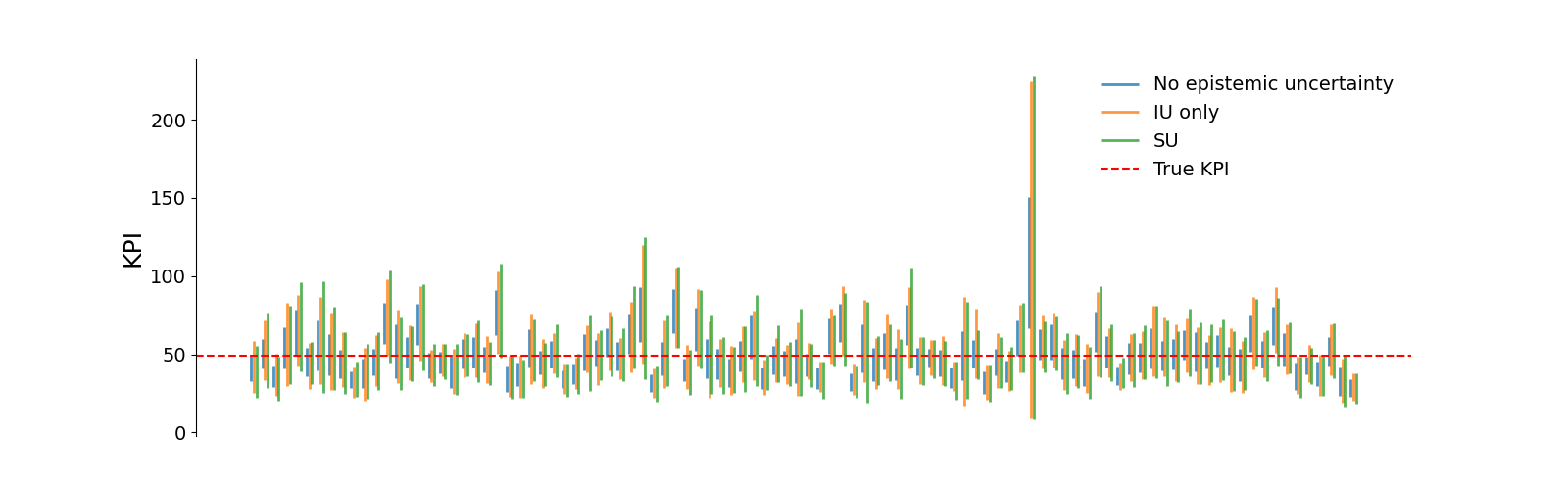}
    \caption{Confidence intervals over 100 macro-replications when accounting for no epistemic uncertainty, only input uncertainty, or submodel uncertainty (SU).}
    \label{fig:first_exp_ci}
\end{figure}

For each case, we conduct $100$ macro-replications.
Within each macro-replication, a training dataset of size $50$ is generated for each submodel by running 50 replications of the simulation model with the true submodels, i.e.,  we obtain 50 observations of $\{X_1, p(X_1), X_2, q(X_2), Y\}$.
We then bootstrap this dataset to generate $B = 100$ training datasets of size 50. For the input models $X_1$ and $X_2$, observations are resampled independently, whereas for the response models $p$ and $q$, the corresponding input-output pairs are resampled jointly. For each bootstrapped dataset, the necessary submodels are estimated and the simulation model is then run for $n = 50$ replications to estimate $\mathbb{E}[Y]$.
On each macro-replication, we construct $90\%$ empirical confidence intervals for $\mathbb{E}[Y]$ via Equation~(\ref{eq:conf}) and calculate their coverages and average widths across macro-replications. The 100 confidence intervals for each case are shown in Figure \ref{fig:first_exp_ci} and their summary statistics are reported in Table~\ref{tab:uq_results}.
As seen in Table~\ref{tab:uq_results}, the coverage and average width of the intervals increase as more sources of uncertainty are considered.

\begin{table}[htbp]
\centering
\caption{Coverage and average confidence interval width under different uncertainty scenarios.}
\label{tab:uq_results}
\begin{tabular}{lccc}
\hline
\textbf{Scenario} & \textbf{Coverage (\%)} & \textbf{Avg. CI Width}\\
\hline
No epistemic uncertainty        & 61.0 & 20.45 $\pm$ 0.14 \\
Input uncertainty only     & 82.0 & 36.14 $\pm$ 0.35\\
Submodel uncertainty    & 87.0 & 38.60 $\pm$ 0.37 \\
\hline
\end{tabular}
\end{table}

\begin{figure}[htbp]
    \centering
    \includegraphics[width=0.8\linewidth]{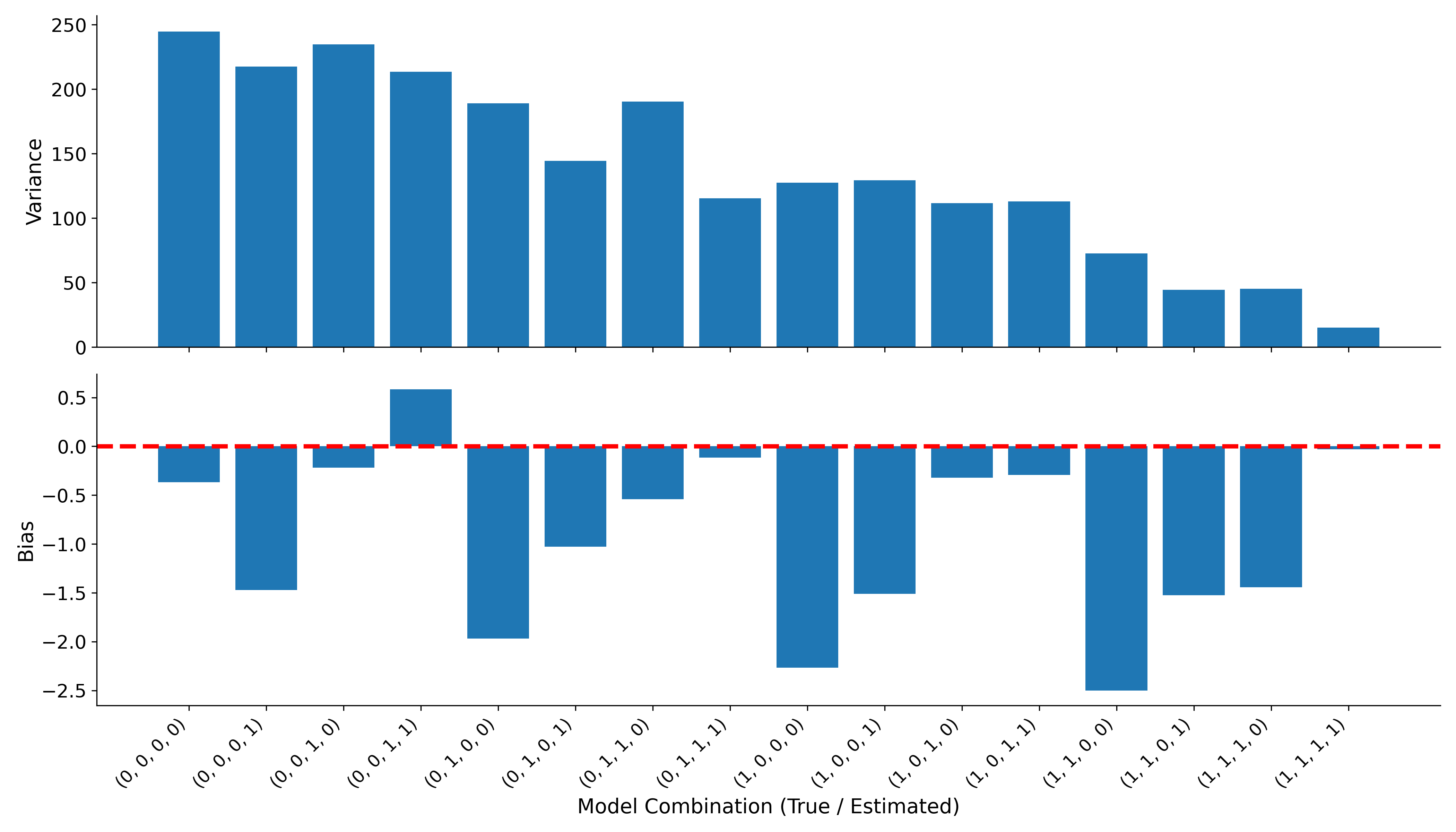}
    \caption{Bias and variance across combinations. The elements in each tuple correspond to $X_1$, $X_2$, $p$, and $q$, respectively; $1$ indicates that the true model is used, and $0$ indicates that the estimated model is used.}

    \label{fig:bias_variance_combinations}
\end{figure}

\begin{figure}[htbp]
    \centering
    \begin{subfigure}[t]{0.48\linewidth}
        \centering
        \includegraphics[width=\linewidth]{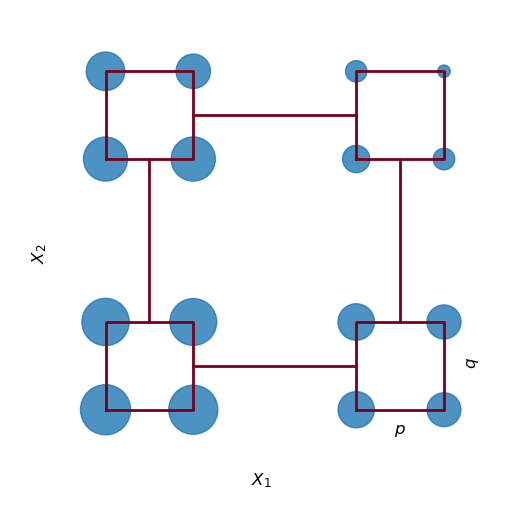}
        \caption{Variance.}
        \label{fig:imp}
    \end{subfigure}
    \hfill
    \begin{subfigure}[t]{0.48\linewidth}
        \centering
        \includegraphics[width=\linewidth]{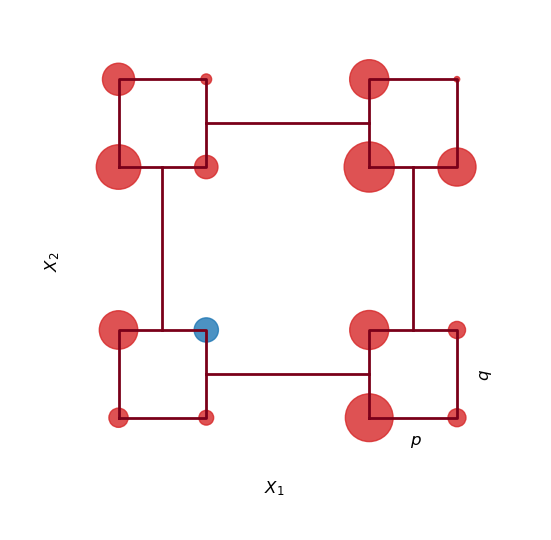}
        \caption{Bias. }
        \label{fig:tmp}
    \end{subfigure}

    \caption{Response grid plots for a $2^4$ full factorial design of the four submodels (true vs estimated). The radius of each circle indicates the relative magnitude. Blue (red) circles show positive (negative) values.}
    \label{fig:rgp_all}
\end{figure}

Because the true data-generating mechanisms are known, we further analyze all $2^4 = 16$ combinations of true and estimated submodels. For each configuration, we generate $500$ independent training datasets of size $50$. When a submodel is estimated in a given configuration, it is fitted using the corresponding training dataset and then embedded in the simulation model and the simulation is run for $n = 500$ replications under that configuration.
Figure \ref{fig:bias_variance_combinations} shows the estimated variance and bias of the resulting estimator for each combination.
Figures~\ref{fig:imp} and \ref{fig:tmp} show response grid plots (RGPs) \citep{barton2025response} where the responses are the variance and the bias, respectively. The plots show that the variance of the estimator decreases when fewer submodels are utilized, and the bias increases when $p(\cdot)$ is estimated. The ANOVA results indicate that $X_1$ ($\text{p-value} = 0.023$) and $p$ ($\text{p-value} \approx 0$) have a statistically significant effect on the output variance, whereas $q$ ($\text{p-value} = 0.04 $) exhibits comparatively weaker evidence of significance and $X_2$ ($\text{p-value} = 0.12$) is not statistically significant.

\begin{figure}[htbp]
    \centering
    \includegraphics[width=0.8\linewidth]{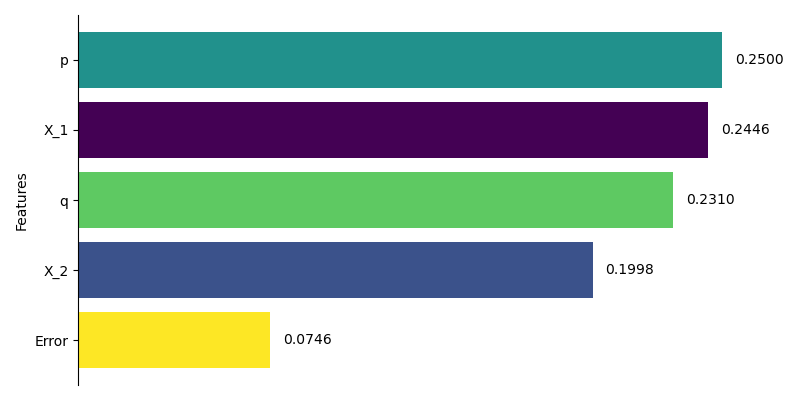}
    \caption{Feature importance plot attributing uncertainty to each of the estimated submodels and aleatoric uncertainty.}
    \label{fig:Feature_importance_plot}
\end{figure}

Figure \ref{fig:Feature_importance_plot} shows the decomposition of the variance across sources of uncertainty when using the bagging approach on one macro replication. The results align with those of the RGP plot and the ANOVA in that $X_1$ and $p$ are identified as contributing the most to the uncertainty. Overall, this suggests that one should prioritize obtaining more data for modeling $X_1$ and training $p$ in order to reduce the epistemic uncertainty.

We also compare the variance of the feature importance scores obtained from the regression tree and the bagging approach. The experiment was conducted over 100 macro-replications. In each macro-replication, $B'= 512$ configurations ($B=8, S=64$) were evaluated using training datasets of size 25 for each submodel. The bagging approach used 50 regression trees. Table~\ref{tab:tree_forest} reports the variance reduction factors (VRFs) across macro-replications and shows that bagging leads to considerable reduction in the variance of the reported importance scores. 

\begin{table}[htbp]
\centering
\caption{Variance reduction factors (VRF) for individual submodels.}
\label{tab:tree_forest}
\renewcommand{\arraystretch}{1.15}
\begin{tabular}{l|cccc}
        \textbf{Submodel} & $\boldsymbol{X_1}$ & $\boldsymbol{X_2}$ & $\boldsymbol{p}$ & $\boldsymbol{q}$ \\
\hline
\textbf{VRF} & 5.4 & 4.7 & 3.6 & 5.6\\
\end{tabular}
\end{table}

\section{Submodel Uncertainty in Digital-Twin Settings}
\label{sec:digial_twin}
A digital twin serves as a virtual counterpart of a physical system and enables forward-looking performance evaluation, robustness analysis, and what-if experimentation under alternative operating scenarios \citep{grieves2017digital}.
A simulation model can be extended into a digital twin by continuously ingesting real-time system information and updating the state, input models, and submodel parameters accordingly. 
In the context of digital twins, a simulation model can be repeatedly initialized using observed system state information and used to forecast KPIs over some future time horizon and support monitoring, planning, and decision making.
To extend our UQ framework to digital-twin settings, we first address how the state of the physical system is observed in an online manner. 
In this setting, our objective is not only to quantify the overall uncertainty in the future KPI given the current state, but also to understand how individual submodels contribute to this uncertainty.

\subsection{Uncertainty Quantification in Digital Twins}
In a digital twin, the state of the physical system is observed at the beginning of an epoch, at which time the simulation model is hot started to predict the KPI at the end of the epoch. 
Let $\mathbf{X}_t$ denote the (observed) state of the system at the start of epoch $t$, and let $\mathbf{S}_t^c$ denote the true subprocesses governing the system during epoch $t$.
Given the state and true subprocesses, the simulation output at the end of epoch $t$ on a given replication \( j = 1, 2, \ldots \) can be expressed as
\[
Y_j(\mathbf{X}_t, \mathbf{S}_t^c)
\equiv \mu(\mathbf{X}_t, \mathbf{S}_t^c)
  + \epsilon_j(\mathbf{X}_t, \mathbf{S}_t^c).
\]

In contrast to the offline simulation setting, the state of the system is now an explicit input to the simulation model that can vary across configurations. As a result, we slightly modify the DOE framework introduced in Section~\ref{sec:uq}. Recall that in the LH designs we assume a common  number of resampled instances for each submodel: $B$. In the digital-twin setting, however, the number of states of interest---whether observed in the past or produced from a future forecast---need not equal $B$. For each state of interest, we propose forming $B'$ design points via a LH design, as described in the previous section, and form a \textit{per-state} dataset as depicted in Table \ref{tab:sampling_states}. For computational reasons, the value of $B'$ may be chosen to be smaller than that in the offline simulation setting. Creating a separate design for each state is motivated by our desire to eliminate the impact of the variability of the system state on the variability of the KPI. We believe that in a digital-twin setting, a decision maker is interested in understanding how much each estimated submodel contributes to the overall uncertainty \textit{across different states}. This is partly because decisions that are informed by the resulting importance scores---such as from which subprocesses more data should be collected---will typically not depend on any one particular state. We therefore propose applying our tree-based method on each state's design and reporting the average importance scores of each submodel across states as indicators of their \textit{aggregate} contributions to epistemic uncertainty. A weighted average can alternatively be used, where the average sum of squares (TSS) in each bagged tree is used as the weight, assigning more importance to states with higher variance.

\subsection{Bias Estimation in Digital Twins}

Recent research has examined the (mis)alignment between digital twins and their physical counterparts \citep{rhodes2023tracking,he2024digital}, where simulation outputs are periodically compared against observed data using statistical tests. We propose a method for quantifying a related form of bias that uses the same data as in the aforementioned design and does not require additional simulation runs.
Suppose that the real system is observed at states
$\{\mathbf{x}_{t_1},\mathbf{x}_{t_2}, \ldots ,\mathbf{x}_{t_N} \}$ with corresponding observed KPIs $\{y(\mathbf{x}_{t_1}, \mathbf{S}_{t_1}^c), y(\mathbf{x}_{t_2}, \mathbf{S}_{t_2}^c), \ldots, y(\mathbf{x}_{t_N}, \mathbf{S}_{t_N}^c) \}$. We define the \textit{state-average bias} to be
$$
\eta_{N}
= \frac{1}{N}
\sum_{i=1}^{N}
\left(
\mathbb{E}\!\left[
Y\!\left(\mathbf{x}_{t_i}, \hat{\mathbf{S}}_{t_i}\right)
\right]
-
\mu\!\left(\mathbf{x}_{t_i}, \mathbf{S}^{c}_{t_i}\right)
\right).
$$
The state-average bias measures the average discrepancy between the simulated and observed KPIs across the set of observed system states. This bias quantifies the systematic misalignment between the digital twin and the physical system that could persist over time if uncorrected. For each configuration in Table \ref{tab:sampling_states}, $b'=1,2,\ldots, B'$, we can calculate a corresponding point estimate of this bias across states:
$$
\hat{\eta}^{\langle b' \rangle}_N = \frac{1}{N} \sum_{i=1}^{N} 
\left(
\frac{1}{n} \sum_{j=1}^{n} 
Y_j\!\left(\mathbf{x}_{t_i}, \hat{\mathbf{S}}_{t_i}^{\langle b' \rangle}\right)
\;-\;
y\!\left(\mathbf{x}_{t_i}, \mathbf{S}^c_{t_i}\right)
\right).
$$
An approximate $100(1-\alpha)\%$ CI for the state-average bias can then be constructed using the empirical quantiles of the
bootstrapped (or resampled) estimates $\hat{\eta}^{\langle 1 \rangle}_N, \hat{\eta}^{\langle 2 \rangle}_N, \dots, \hat{\eta}^{\langle B' \rangle}_N$ as
\[
  \left[
Q_{\alpha/2} (\hat{\eta}_N),
\;
Q_{1-\alpha/2} (\hat{\eta}_N)
\right],
\]
where \( Q_p(\hat{\eta}_N) \) denotes the empirical $p$-quantile.

\begin{table}[h!]
    \caption{Sampling framework for submodel uncertainty in digital-twin setting.}
    \centering
    \renewcommand{\arraystretch}{1.5}
    \begin{tabular}{c|c|c}
        \textbf{State} 
        & \textbf{Submodels} 
        & \textbf{Output} \\
        \hline

        \multirow{4}{*}{$\mathbf{X}_{t}$}
        & \multirow{4}{*}{$\hat{\mathbf{S}}_{t}^{\langle 1 \rangle}$}
        & $Y_{t,1}^{\langle 1 \rangle}
           = Y_1(\mathbf{X}_{t},
               \hat{\mathbf{S}}_{t}^{\langle 1 \rangle})$
        \\
        & & $Y_{t,2}^{\langle 1 \rangle}
           = Y_2(\mathbf{X}_{t},
               \hat{\mathbf{S}}_{t}^{\langle 1 \rangle})$
        \\
        & & $\vdots$ \\
        & & $Y_{t,n}^{\langle 1 \rangle}
           = Y_n(\mathbf{X}_{t},
               \hat{\mathbf{S}}_{t}^{\langle 1 \rangle})$
        \\
        \hline

        \multirow{4}{*}{$\mathbf{X}_{t}$}
        & \multirow{4}{*}{$\hat{\mathbf{S}}_{t}^{\langle 2 \rangle}$}
        & $Y_{t,1}^{\langle 2 \rangle}
           = Y_1(\mathbf{X}_{t},
               \hat{\mathbf{S}}_{t}^{\langle 2 \rangle})$
        \\
        & & $Y_{t,2}^{\langle 2 \rangle}
           = Y_2(\mathbf{X}_{t},
               \hat{\mathbf{S}}_{t}^{\langle 2 \rangle})$
        \\
        & & $\vdots$ \\
        & & $Y_{t,n}^{\langle 2 \rangle}
           = Y_n(\mathbf{X}_{t},
               \hat{\mathbf{S}}_{t}^{\langle 2 \rangle})$
        \\
        \hline

        $\vdots$ & $\vdots$ & $\vdots$ \\
        \hline

        \multirow{4}{*}{$\mathbf{X}_{t}$}
        & \multirow{4}{*}{$\hat{\mathbf{S}}_{t}^{\langle B' \rangle}$}
        & $Y_{t,1}^{\langle B' \rangle}
           = Y_1(\mathbf{X}_{t},
               \hat{\mathbf{S}}_{t}^{\langle B' \rangle})$
        \\
        & & $Y_{t,2}^{\langle B' \rangle}
           = Y_2(\mathbf{X}_{t},
               \hat{\mathbf{S}}_{t}^{\langle B' \rangle})$
        \\
        & & $\vdots$ \\
        & & $Y_{t,n}^{\langle B' \rangle}
           = Y_n(\mathbf{X}_{t},
               \hat{\mathbf{S}}_{t}^{\langle B' \rangle})$
        \\
    \end{tabular}
    \label{tab:sampling_states}
\end{table}

\subsection{Numerical Example}
\label{sec:example_2}
To illustrate the effects of submodel uncertainty in the digital-twin setting, we consider a simulation model of a contact center with two contact groups and three expert groups. Contacts from each group are assumed to arrive according to two independent nonhomogeneous Poisson processes with piecewise-constant arrival rate functions that vary by hour. For each contact group, patience times and service times are jointly distributed with Gamma marginal distributions coupled through a Gaussian copula.
The KPI of interest is the expected average waiting time of contacts in the second contact group. Each expert group has two experts. The first expert group serves only the first contact group, the second expert group serves only the second contact group, and the third expert group serves both contact groups with priority given to the first contact group.
 
A routing mechanism dynamically assigns contacts to experts based on system conditions and is triggered when one of two scenarios occurs: If a contact arrives to find their contact group queue is empty and one or more compatible experts are available, the contact is immediately assigned to one of the available experts. Alternatively, if an expert becomes available when queues are present, a waiting contact is selected to be serviced by that expert. We consider two distinct routing processes: a contact-triggered routing process, denoted by $\mathbf{S}^c_C$, and an expert-triggered routing process, denoted by $\mathbf{S}^c_E$. Both $\mathbf{S}^c_C$ and $\mathbf{S}^c_E$ are rules-based functions with non-trivial decision rules. For the frequentist case, we use maximum likelihood estimators (MLEs) for the input submodels and logistic regression models for the routing submodels. For the Bayesian case, we use non-informative priors for the input submodels---which yields the same estimates as the MLEs---and for the routing submodels, we train two Bayesian neural networks, each with one hidden layer with 128 nodes.

\begin{figure}[htbp]
    \centering
    \begin{subfigure}[t]{0.83\linewidth}
        \centering
        \includegraphics[width=\linewidth]{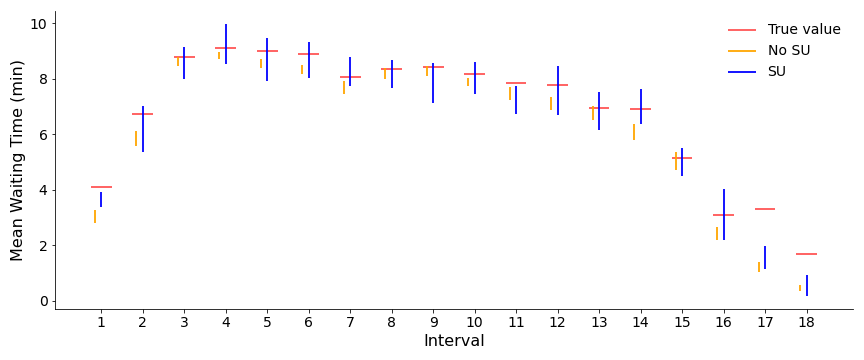}
        \caption{Frequentist submodels.}
        \label{fig:SU_second_exp_nonbayes}
    \end{subfigure}
    \hfill
    \begin{subfigure}[t]{0.83\linewidth}
        \centering
        \includegraphics[width=\linewidth]{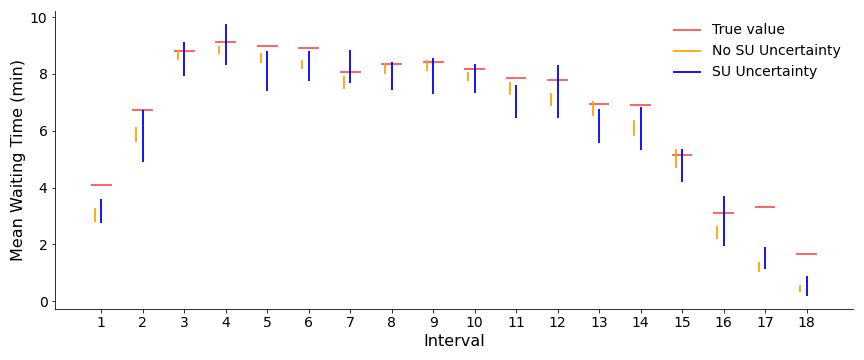}
        \caption{Bayesian submodels.}
        \label{fig:SU_second_exp_bayes}
    \end{subfigure}

    \caption{90\% confidence and credible intervals when considering submodel uncertainty and not considering submodel uncertainty for (a) frequentist and (b) Bayesian submodels.}
    \label{fig:SU_second_exp}
\end{figure}

\begin{figure}[htbp]
    \centering
    \begin{subfigure}[t]{0.48\linewidth}
        \centering
        \includegraphics[width=\linewidth]{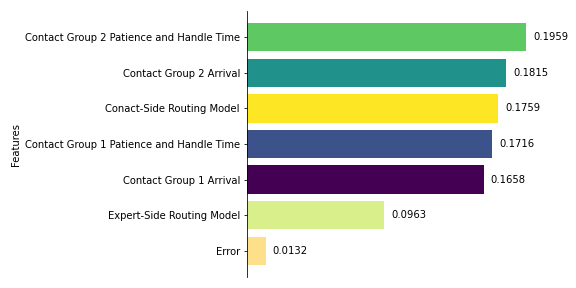}
        \caption{Frequentist submodels.}
        \label{fig:importance_second_exp_nonbayes}
    \end{subfigure}
    \hfill
    \begin{subfigure}[t]{0.48\linewidth}
        \centering
        \includegraphics[width=\linewidth]{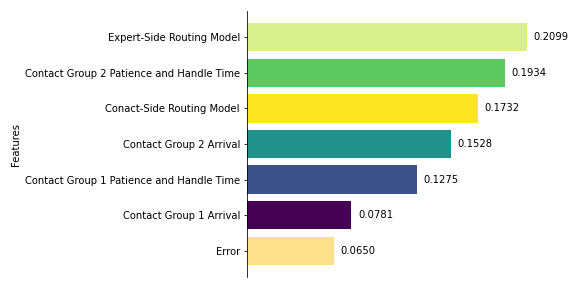}
        \caption{Bayesian submodels.}
        \label{fig:importance_second_exp_bayes}
    \end{subfigure}

    \caption{Importance scores for the submodels in the second experiment for (a) frequentist and (b) Bayesian formulations.}
    \label{fig:importance_plot_second_exp}
\end{figure}

The simulation model is run for a single nine-hour day, and the state of the system is recorded at each 30-minute interval for a total of 18 epochs. At the end of the day, data on customer arrivals, patience times, handle times, and routing decisions are collected and used to construct the estimated submodels.
In the case without submodel uncertainty, the simulation is run for 100 replications for each of the observed system states. In the presence of submodel uncertainty, 180 configurations are selected, where for each configuration we use one of the five resampled instances of each submodel as denoted by the Latin hypercube (LH) design. Since we have observed 18 states, we pick 2 stacks each with 5 configurations of submodels for each state.
To obtain the true KPI value for each state, an additional 1{,}000 replications are performed and the results treated as the ground-truth benchmark.
Figure~\ref{fig:SU_second_exp} reports 90\% confidence and credible intervals for the mean waiting-time KPI of the second contact group at the end of each epoch under the two experimental settings for both frequentist and Bayesian settings. As observed, the width of the intervals increases as submodel uncertainty is incorporated. For each observed system state, we compute importance scores using the bagging approach with 40 trees and average these scores across all states to obtain an overall importance measure for both the frequentist and Bayesian settings, as shown in Figure~\ref{fig:importance_plot_second_exp}. The second contact group's arrival and patience-and-handle time submodels contribute the most to the overall uncertainty. Therefore one should prioritize enhancing the second contact group's input modeling in the frequentist setting, for instance, by collecting more data. In the Bayesian setting, the expert-side routing submodel contributes the most to the overall uncertainty, thus collecting more training data or considering a more flexible ML model could potentially reduce the epistemic uncertainty. Calculating the overall bias yields CIs of $(-0.33, -0.12)$ and $(-0.61, -0.38)$ for the frequentist and Bayesian settings, respectively. These results indicate that the use of estimated submodels has led to a potential misalignment in KPI estimation. Whether this degree of state-average bias is practically significant will depend on the business purpose.

\section{Conclusion}
\label{sec:conclusion}
This paper develops a general framework for quantifying uncertainty that arises from embedding submodels within stochastic simulation models. We employ bootstrapping and Bayesian model averaging to construct quantile-based confidence and credible intervals that capture both epistemic and aleatoric uncertainty. To understand the drivers of this uncertainty, we use regression-tree-based variable-importance measures to identify the most influential submodels. 
We further show how the proposed framework can be extended to digital-twin settings with time-varying submodels. Numerical experiments demonstrate that conventional methods may substantially underestimate uncertainty, whereas accounting for submodel uncertainty yields more reliable inference. Future work can investigate the role of uncertainty quantification in model calibration and active learning approaches for dynamically training and updating submodels in digital-twin settings.

\section*{Acknowledgement}
{This work was supported by the Intuit University Collaboration Program. We also thank Dusan Bosnjakovic for his support throughout the project.}






\footnotesize

\bibliographystyle{plain}

\bibliography{template}

\end{document}